\documentclass[12pt]{article}

\usepackage{amsfonts}
\usepackage{amsmath}
\usepackage[dvipdfmx]{hyperref}
\usepackage{cite}
\usepackage{epsfig}
\usepackage{latexsym}
\usepackage{paralist}
\usepackage{fancyhdr}
\usepackage{graphicx}
\numberwithin{equation}{section}
\usepackage[vcentermath]{youngtab}
\usepackage{young}
\usepackage{ytableau}
\usepackage{etex}
\usepackage{braket}
\usepackage{float}
\usepackage{slashed}

\def\be{\begin{equation}}
\def\ee{\end{equation}}
\def\bea{\begin{eqnarray}}
\def\eea{\end{eqnarray}}

\renewcommand{\thefootnote}{\fnsymbol{footnote}}

\begin{document}

\hfuzz=100pt
\title{{\Large \bf{Classical equation of motion and\\ Anomalous dimensions at leading order}}}
\date{}
\author{ Keita Nii$^a$\footnote{keitanii@hri.res.in}
}
\date{\today}

\maketitle

\thispagestyle{fancy}
\cfoot{}
\renewcommand{\headrulewidth}{0.0pt}

\vspace*{-1cm}
\begin{center}
$^{a}${{\it Harish-Chandra Research Institute }}
\\ {{\it Chhatnag Road, Jhusi, Allahabad 211019, India}}

\end{center}

\begin{abstract}
Motivated by a recent paper by Rychkov-Tan \cite{Rychkov:2015naa}, we calculate the anomalous dimensions of the composite operators at the leading order in various models including a $\phi^3$-theory in $(6-\epsilon)$ dimensions. The method presented here relies only on the classical equation of motion and the conformal symmetry.
In case that only the leading expressions of the critical exponents are of interest, it is sufficient to reduce the multiplet recombination discussed in \cite{Rychkov:2015naa} to the classical equation of motion. We claim that in many cases the use of the classical equations of motion and the CFT constraint on two- and three-point functions completely determine the leading behavior of the anomalous dimensions at the Wilson-Fisher fixed point without any input of the Feynman diagrammatic calculation. The method developed here is closely related to the one presented in \cite{Rychkov:2015naa} but based on a more perturbative point of view.
\end{abstract}

\renewcommand{\thefootnote}{\arabic{footnote}}
\setcounter{footnote}{0}

\newpage
\clearpage

\section{Introduction}
Recently, CFT techniques have been extensively used to extract the critical behavior of the various quantum field theories in dimensions larger than two. Among them, the numerical conformal bootstrap is one of the most successful tools \cite{ElShowk:2012ht}. In a paper by Rychkov-Tan \cite{Rychkov:2015naa}, the multiplet recombination was used to gain the (irrelevant) critical exponents of the Wilson-Fisher fixed point of the $\phi^4$-theory in $(4-\epsilon)$ dimensions. They assumed that the $\phi^3$ operator should be a descendant state of the elementary $\phi$ field and thus the quantum dimension of $\phi^3$ is related to the one of $\phi$ and then the field $\phi^3$ should appear in the OPE as the descendant. Although, at the Gaussian fixed point, these two operators are both primary but for the WF fixed point, $\phi^3$ is a member of the $\phi$ conformal multiplet. They called this situation ``multiplet recombination''.
They have studied the two-point functions and the three-point functions by using the multiplet recombination and the OPE and they have found the anomalous dimensions of the composite operators $\phi^n$ at the leading order in $\epsilon$ without any diagrammatic calculation. 

The method developed in \cite{Rychkov:2015naa} was immediately applied to other types of the ``WF fixed point'', for example, the $\phi^6$-theory in $(3-\epsilon)$ dimensions \cite{Basu:2015gpa} and the Gross-Neveu model in $(2+\epsilon)$ dimensions \cite{Raju:2015fza,Ghosh:2015opa}, where the anomalous dimensions of various composite operators are derived at the leading order in $\epsilon$. In all these examples, it is important that the anomalous dimension of the elementary field is $O(\epsilon^2)$ and the ones of the composite operators are $O(\epsilon)$. Then the naive application of the method \cite{Rychkov:2015naa} to the $\phi^3$-theory in $(6-\epsilon)$ dimensions did not work due to the fact that in the six-dimensional $\phi^3$-theory the wave function renormalization of the field $\phi$ starts with the one-loop graph.  
This method is quite generic and does not require the perturbative treatment, which implies that potentially one can predict the non-perturbative results beyond the leading order approximation. However the consideration of only the two- and three-point functions and the OPE give the leading expression for the critical exponents at $O(\epsilon)$ or $O(\epsilon^2)$.\footnote{The next leading order of the anomalous dimensions was studied in \cite{Sen:2015doa} by using the CFT techniques and the unitarity.}

In this paper, we will study and re-organize the Rychkov-Tan method from the more perturbative and more Lagrangian-based point of view in order to adapt the method also to the six-dimensional $\phi^3$-theory. The similar analysis was already done in \cite{Anselmi:1998bh, Belitsky:2007jp} and recently in \cite{Giombi:2016hkj}. The main point of the re-organized method is the use of the Schwinger-Dyson equation without contact terms which do not contribute to the discussion here. This is not the same as the multiplet recombination used in \cite{Rychkov:2015naa} since the multiplet recombination generally can not be derived from the Lagrangian-based approach. However, for the leading order calculation, these are effectively equivalent. 
The Schwinger-Dyson equation for the scalar field theory is schematically written as
\begin{align}
\braket{\Box_x \phi(x)  O_1(x_1)  O_2 (x_2) \cdots } = \braket{\frac{g}{3!}\phi^3(x) O_1(x_1)  O_2 (x_2) \cdots  },
\end{align}
where we considered the $\phi^4$-theory and neglected the contact terms. This equation should be quantum-mechanically regarded as the renormalized one.
However, when we estimate the right-hand side at $O(g)$, the tree-level evaluation of the correlation function suffices for our purpose and then all the quantities can be reduced to the bare quantities. The Schwinger-Dyson equation reduces to the classical equation of motion. 
If we restrict our attention to the conformal field theory and to the two- or three-point functions of the conformal primary operators, the left-hand side can be fixed up to the constant coefficient. Due to the derivative $\Box_x$, the left-hand side is proportional to the anomalous dimensions of the primary operators. Equating both sides, we will find the values of the anomalous dimensions as the function of the coupling. 

For the Wilson-Fisher fixed point we need to know the critical coupling $g_*$. Our claim in this paper is that considering the two- and three-point functions and employing the Schwinger-Dyson equation (classical equation of motion), we can decide the value of $g_*$ without any input from the loop calculation. The use of the Schwinger-Dyson equation for the calculation of the anomalous dimensions is very reminiscent of the story that the renormalized equation of motion was used for relating the critical exponents and for reducing the number of the independent exponents \cite{Amit:1984ms,Brezin:1974zr}. 

We will study the $\phi^3$-theory in $(6 - \epsilon)$ dimensions and find the leading critical exponent at $O(\epsilon)$ without any input from the Feynman diagrammatic calculation. The similar analysis was recently carried out in \cite{Giombi:2016hkj}, where however the critical coupling $g_*$ is decided from the $\beta$ function derived from the perturbative calculation. The calculation in this paper is very respecting the method by Rychkov-Tan \cite{Rychkov:2015naa} in a sense that all the calculations include no Feynman diagrammatic calculation and use the general forms of the two- and three-point functions of the conformal primary operators although our method is more closer to the perturbative method than the one by Rychkov-Tan \cite{Rychkov:2015naa}.
As the result, when we study the higher order behavior of the critical exponents, we will inevitably need the perturbative calculation of the Feynman graphs but the use of the equation of motion will reduce the complexity of the perturbative calculation.

The organization of the paper is as follows. In first two sections we will study the $\phi^6$-theory in $(3-\epsilon)$ dimensions and the $\phi^4$-theory in $(4-\epsilon)$ dimensions respectively. These two sections serve as  an explanation of our method.
In Section 4, we will consider the $\phi^3$-theory in $(6-\epsilon)$ dimensions.
In Section 5, we generalize the result of Section 4 by introducing the additional $O(N)$ scalar fields and modifying the interaction.
 In Section 6, we summarize the results and discuss the potential future directions.
 
\section{$\phi^6$-theory in $(3-\epsilon)$ dimensions} 
We will first consider the $\phi^6$-theory in $d=3- \epsilon$ dimensions as an illustration of our strategy. This theory has an analog of the Wilson-Fisher fixed point in $(4-\epsilon)$ dimensions. The perturbative treatment can be found, for example, in \cite{Pisarski:1982vz,Hager:2002uq,Lewis:1978zz} and the conformal method using the multiplet recombination was performed in \cite{Basu:2015gpa}. In the following calculation, we assume that all the $\phi^n$ operators except for $n=5$ are the primary operators and that the conformal symmetry appears at the ``Wilson-Fisher fixed point''.

\subsubsection*{Set-up}
The action of the $\phi^6$-theory is
\begin{align}
S=\int d^d x \,  \left( \frac{1}{2} \partial \phi^2 +\frac{g \mu ^{2 \epsilon}}{6!} \phi^6   \right),~~~d=3-\epsilon
\end{align}
and we will maintain the coupling constant in a dimensionless one. We are only interested in the leading order calculations and then the $\mu^{2\epsilon}$ factor does not play any role. Therefore we will omit this factor in the following discussion. For the purpose of the perturbative (Feynman diagrammatic) calculation, we have to include the other renormalizable terms, but in our calculation these are not required.
The scaling dimensions for the field $\phi$ and the composite operators $\phi^n$ are defined as
\begin{align}
\Delta_1  &:= \Delta_{\phi} =\frac{1-\epsilon}{2} +\gamma_1 \\
\Delta_n  &:= \Delta_{\phi^n} =n\left( \frac{1-\epsilon}{2}\right) +\gamma_n.
\end{align}
For the calculation of the lowest order anomalous dimension, the multiplet recombination employed in \cite{Rychkov:2015naa} is nothing but the classical equation of motion
\begin{align}
\Box \phi =\frac{g \mu^{2\epsilon} }{5!} \phi^5.
\end{align}
in our approach.
In the paper \cite{Rychkov:2015naa}, the multiplet recombination relation is defined as $\Box [\phi]_R  = \alpha(\epsilon) [\phi^5]_R$, where $[\cdots]_R$ means the renormalized operator at the Wilson-Fisher fixed point and $\alpha(\epsilon)$ is a certain function which becomes zero at the $\epsilon \rightarrow 0$ limit. Although this is a more generic operator identity and would be necessary for the higher order calculation, but in the lowest order calculation of the anomalous dimensions, the tree-level relation, namely the classical equation of motion is adequate.

\subsubsection*{Two-point function}
As a first step of determining the anomalous dimension from the conformal symmetry and the classical equation of motion, we will study the two-point function of the elementary field $\phi$. This step is completely the same as \cite{Rychkov:2015naa,Basu:2015gpa,Giombi:2016hkj}. The two-point function is given by
\begin{align}
\braket{\phi(x)  \phi(y)} \overset{\mathrm{tree-level}}{=} \frac{1}{4 \pi} \frac{1}{|x-y|}\\
\braket{\phi(x)  \phi(y)} =c \, |x-y|^{-2 \Delta_1},
\end{align}
where the tree-level value of $c$ is $1/4\pi$.
Correctly speaking, in the second equation, the operator should be a renormalized one $[\phi]_R$. In the following calculation we will multiply the above function by the differential operator $\Box:=\partial_\mu \partial^\mu$ twice and use the equation of motion $\Box \phi =\frac{g \mu^{2\epsilon} }{5!} \phi^5$. Since the equation of motion introduces the coupling dependence and all the correlation functions will be evaluated at the tree-level, we can replace all the renormalized operator with the tree-level (bare) ones. Therefore we can neglect the wave function renormalization in the following discussion. For the same reason, the renormalization factor of the coupling constant can be dropped.

As mentioned above, taking the derivatives of the above two-point function first by $\Box_x$, we have 
\begin{align}
\braket{ \Box_x \phi(x)  \phi(y)} &= c \Box_x |x-y|^{-2 \Delta_1}  \nonumber \\
&= 2c  \Delta_1(2\Delta_1 +2-d)  |x-y|^{-2\Delta_1- 2} \nonumber \\
&\sim   \frac{\gamma_1}{2 \pi} |x-y|^{-5}, 
\end{align}
where in the last line we have evaluated this at the lowest level in the coupling constant and the epsilon parameter. Since we eventually find the critical coupling $g_* =g(\epsilon)$, neglecting the $\epsilon$-dependence above will suffice for our purpose. We can alternatively calculate the above correlation functions by using the classical equation of motion as
\begin{align}
\left. \braket{ \Box \phi(x) \phi(y)}\right|_{\mathrm{lowest}} &= \left. \frac{g \mu^{2\epsilon}}{5!} \ \braket{\phi^5(x)  \phi(y)} \right|_{\mathrm{lowest}}=0,
\end{align}
which means $\gamma_1=O(g^2)$. Next we will further apply the differential operator $\Box_y$ in addition to $\Box_x$,
\begin{align}
\braket{ \Box \phi(x) \Box \phi(y)} &= c \Box_x \Box_y |x-y|^{-2 \Delta_1} \nonumber  \\
&= 2c  \Delta_1(2\Delta_1 +2) (2\Delta_1 +2-d) (2\Delta_1+4-d) |x-y|^{-2\Delta_1 -4} \nonumber  \\
&\sim \frac{3}{\pi} \gamma_1 |x-y|^{-5}.
\end{align}
We can again evaluate the two-point function of $\Box \phi$ by using the classical equation of motion as follows. At $O(g^2)$, all the calculations are carried out in a tree-level approximation.
\begin{align}
\left. \braket{ \Box \phi(x) \Box \phi(y)} \right|_{O(g^2)} &= \left( \frac{g \mu^{2\epsilon}}{5!} \right)^2 \braket{\phi^5(x)  \phi^5(y)} \nonumber \\
&=\frac{g^2}{5!} \frac{1}{(4 \pi)^5} |x-y|^{-5}
\end{align}
In the calculation above we again retain the lowest order. Comparing these two results, we obtain
\begin{align}
\gamma_1 =\frac{1}{3 \cdot 4^5 \cdot 5! \, \pi^4} g^2.
\end{align}
As the calculation above shows, all the renormalized quantities $[\phi]_R$ and $g_R$ can be replaced by the bare quantity for the lowest order calculation.
Notice that here we are using the conformal properties of the primary operators so we should tune the coupling constant $g$ to a special value to realize the Wilson-Fisher fixed point. Usually we can calculate $g_* =g(\epsilon)$ by the Feynman diagrammatic approach, but we will alternatively find this critical value by considering the three-point functions without any diagrammatic input, which is based on the philosophy of \cite{Rychkov:2015naa} (and for the extended works see \cite{Basu:2015gpa, Raju:2015fza,Ghosh:2015opa}).

\subsubsection*{Three-point functions}
Next we will study the three-point functions. The conformal symmetry restricts the coordinate dependence of the three-point function of the (quasi-)primary operators as
\begin{align}
\braket{\phi(x) \phi^n (y) \phi^{n+1} (z)  } &= C_{1,n,n+1} |x-y|^{\Delta_{n+1}-\Delta_{1}-\Delta_{n}} |y-z|^{\Delta_{1}-\Delta_{n}-\Delta_{n+1}} |x-z|^{\Delta_{n}-\Delta_{1}-\Delta_{n+1}}, 
\end{align}
where we assume $n \ge 2$ and the tree-level value of $C_{1,n,n+1}$ is given by
\begin{align}
C_{1,n,n+1}^{\mathrm{tree}} &=(n+1)! \frac{1}{(4 \pi)^{n+1}}.
\end{align}
These three-point functions actually include the non-primary operator for $n=4,5$ since $\phi^5$ is assumed to be a descendant at the conformal fixed point we want to analyze. Then for $n=4,5$ the right-hand side is not correct and includes the various terms like
\begin{align}
\sum_{a,b} C_{a,b}|x-y|^a|y-z|^b|z-x|^{-a-b-\Delta_{1}-\Delta_{n}-\Delta_{n+1}}.
\end{align}
However, since we are now interested in the lowest order values of the anomalous dimensions, the above restricted form is adequate even for $n=4$ and $5$. This will be proven in the last paragraph of this section.

In the same way as the two-point function, we apply the differential operator $\Box_x$ and evaluate it at the lowest-level,
\begin{align}
\braket{ \Box_x \phi(x) \phi^n (y) \phi^{n+1} (z)  } &\sim C_{1,n,n+1} \gamma_1 |y-z|^{-n} |x-z|^{-3} \nonumber \\
& \qquad -C_{1,n,n+1} (\gamma_1+\gamma_{n}-\gamma_{n+1}) |x-y|^{-2} |y-z|^{-n+2} |x-z|^{-3}.
\end{align}
The terms proportional to $\gamma_1$ will start with $O(g^2)$, then these do not contribute to the following discussion with $O(g)$. Using the classical equation of motion, we find
\begin{align}
\braket{ \Box_x \phi(x) \phi^n (y) \phi^{n+1} (z)  } &=\frac{g \mu^{2\epsilon}}{12 (4 \pi)^{n+3}} n(n-1)(n+1)! |x-y|^{-2} |y-z|^{-n+2} |x-z|^{-3}.
\end{align}
Equating these two results we obtain the recursion relation
\begin{align}
\gamma_{n+1} -\gamma_{n}=\frac{n(n-1)}{12 (4 \pi)^2}g +O(g^2) ~~~\mbox{for}~ n \ge 2,
\end{align}
where we have used the tree-level value of $C_{1,n,n+1}$.

Next we will consider the following three-point function
\begin{align}
\braket{\phi(x) \phi(y)  \phi^2(z)  } &=C_{1,1,2} |x-y|^{\Delta_2 -2\Delta_1} |y-z|^{-\Delta_2} |x-z|^{-\Delta_2},  
\end{align}
where the tree-level value of the coefficient $C_{1,1,2}$ is given by 
\begin{align}
C_{1,1,2}^{\mathrm{tree}}=\frac{1}{8 \pi^2}.
\end{align}
First, we take the derivative by $\Box_x$ and evaluate it at the lowest order.
\begin{align}
\braket{\Box_x \phi(x) \phi(y)  \phi^2(z) } &=C_{1,1,2} (2 \Delta_1 -\Delta_2) (2\Delta_1 +2-d) |x-y|^{\Delta_2 -2\Delta_1-2}|y-z|^{-\Delta_2}|x-z|^{-\Delta_2}  \nonumber  \\ 
& \qquad +C_{1,1,2} \Delta_2 (2 \Delta_1+2-d) |x-y|^{\Delta_2 -2 \Delta_1} |y-z|^{-\Delta_2} |x-z|^{-\Delta_2-2}   \nonumber \\ 
&\qquad \quad -C_{1,1,2} \Delta_2 (2\Delta_1 -\Delta_2)  |x-y|^{\Delta_2 -2\Delta_1-2}|y-z|^{-\Delta_2+2}|x-z|^{-\Delta_2-2}  \nonumber \\
&\sim 2C_{1,1,2} \gamma_1 |y-z|^{-1}|x-z|^{-3}-C_{1,1,2} (2\gamma_1 -\gamma_2) |x-y|^{-2}|y-z| |x-z|^{-3}
\end{align}
If we use the equation of motion and evaluate the left-hand side at the lowest order, the result is vanishing, which means that the non-vanishing terms will start with $O(g^2)$. This implies that the anomalous dimension $\gamma_2$ will start at $O(g^2)$. We further take a derivative by $\Box_y$ and we find
\begin{align}
\braket{\Box_x \phi(x) \Box_y \phi(y)  \phi^2(z) } \sim -2 C_{1,1,2} (2\gamma_1-\gamma_2 )|x-y|^{-4} |y-z|^{-1} |x-z|^{-1} +O(g^3).
\end{align}
By using again the equation of motion, the left-hand side becomes
\begin{align}
\braket{\Box_x \phi(x) \Box_y \phi(y)  \phi^2(z) } &= \left( \frac{g}{5!} \right)^2 \braket{\phi^5 (x) \phi^5 (y) \phi^2} \nonumber \\
&\sim \frac{g^2}{12 (4 \pi)^6} |x-y|^{-4} |y-z|^{-1} |x-z|^{-1} 
\end{align}
Comparing these two results we find 
\begin{align}
2\gamma_1-  \gamma_2 =-\frac{g^2}{3 \cdot 4^6 \pi^4}
\end{align}
and the recursion relation above can be solved by noticing that $\gamma_2 =0\cdot g^1 +O(g^2)$,
\begin{align}
\gamma_n= \frac{g}{3^2 4^3 \pi^2}n(n-1)(n-2) +O(g^2),~~~\mbox{for}~~ n \ge 2
\end{align}
Note again that the above derivation uses the conformal properties so we need to find the critical coupling $g_*$. This will be given in the following.

\subsubsection*{Determination of $g_*$}
We should notice that $\Delta_5$ can be represented in two ways,
\begin{align}
\Delta_5 &=\Delta_1 +2 =\frac{5 -\epsilon}{2} +\gamma_1 \\
&=5\left(  \frac{1-\epsilon}{2} \right) +\gamma_5
\end{align}
Then we get the relation
\begin{align}
\gamma_5-\gamma_1 =2\epsilon.
\end{align}
At $O(g)$, this means $\gamma_5 =2 \epsilon$, which fixes the value of $g_*$ as
\begin{align}
g_*=\frac{2^5 \cdot 3 \pi^2 }{5} \epsilon
\end{align}
Inserting this to the anomalous dimensions obtained above, we find
\begin{gather}
\gamma_1=\frac{1}{1000} \epsilon^2+O(\epsilon^3),~~\gamma_2=\frac{3}{100} \epsilon^2 +O(\epsilon^3)\\
\gamma_{n \ge 3} =\frac{1}{30} n(n-1)(n-2)\epsilon+O(\epsilon^2).
\end{gather} 
These results are consistent with the perturbative calculation \cite{Hager:2002uq, Lewis:1978zz} and the conformal method \cite{Basu:2015gpa}. In the paper \cite{Basu:2015gpa}, the value of $\gamma_2$ is not derived. 

\subsubsection*{OPE coefficients}
Next we will consider the OPE coefficients which are vanishing in the limit $g \rightarrow 0$.
Let us consider the following three-point function
\begin{align}
\braket{\phi(x)  \phi(y)  \phi^4 (z)} &=C_{1,1,4} |x-y|^{\Delta_4-2\Delta_1} |y-z|^{-\Delta_4}  |x-z|^{-\Delta_4} \nonumber \\
\braket{ \Box_x \phi(x)  \phi(y)  \phi^4 (z)} & \sim 2 C_{1,1,4} |x-y|^{-1} |x-z|^{-4}. 
\end{align}
Again, by the use of the equation of motion, the left-hand side becomes
\begin{align}
\braket{ \Box_x \phi(x)  \phi(y)  \phi^4 (z)} &=\frac{g}{5!} \braket{\phi^5(x) \phi(y) \phi^4(z)} \nonumber \\
&\sim \frac{g}{(4 \pi)^5} |x-y|^{-1} |x-z|^{-4}.
\end{align}
Then we obtain the lowest order coefficient
\begin{align}
\left. C_{1,1,4} \right|_{O(g)} =\frac{g}{2(4 \pi)^5}.  
\end{align}
In a similar way, considering the three-point function $\braket{\phi (x) \phi(y) \phi^6(z)}$, we obtain
\begin{align}
\left. C_{1,1,6} \right|_{O(g)} =\frac{g}{(4 \pi)^6}.
\end{align}
%

\subsubsection*{Justification of the form of the 3-pt function for $n=3,4$}
In the above calculation, we assume the following form of the three-point function. This is correct for the (quasi-)primary operators but not for their descendants. For the descendant fields, we generically have the different terms. Here we will justify that in the lowest order calculation it is enough to restrict the form of the three-point function to the following form. 
\begin{align}
\braket{\phi(x) \phi^n (y) \phi^{n+1} (z)  } &= C_{1,n,n+1} |x-y|^{\Delta_{n+1}-\Delta_{1}-\Delta_{n}} |y-z|^{\Delta_{1}-\Delta_{n}-\Delta_{n+1}} |x-z|^{\Delta_{n}-\Delta_{1}-\Delta_{n+1}}, 
\end{align}

Let us first consider the three-point function,
\begin{align}
\braket{\phi(x) \phi^4(y)  \phi^5(z)}.
\end{align}
Since $\phi^5(z)$ is a descendant of a single $\phi(z)$ field, this is proportional to 
\begin{align}
\frac{1}{g} \braket{\phi (x) \phi^4(y) \Box_z \phi(z) }.
\end{align}
This correlation function is constrained by the conformal symmetry as follows.
\begin{align}
\braket{\phi (x) \phi^4(y) \Box_z \phi(z) } &=C_{1,4,1} \Box_z \left( |x-y|^{-\Delta_4} |y-z|^{-\Delta_4} |x-z|^{\Delta_4-2 \Delta_1}\right)  \nonumber \\
&=C_{1,4,1} \Delta_4 (2\Delta_1 +2-d) |x-y|^{-\Delta_4} |y-z|^{-\Delta_4-2} |x-z|^{\Delta_4-2\Delta_1} \nonumber \\
&\quad +C_{1,4,1} (2\Delta_1 -\Delta_4) (2\Delta_1 +2-d) |x-y|^{-\Delta_4} |y-z|^{-\Delta_4} |x-z|^{\Delta_4 -2\Delta_1-2} \nonumber  \\
&\qquad -C_{1,4,1} \Delta_4 (2 \Delta_1 -\Delta_4) |x-y|^{-\Delta_4 +2} |y-z|^{-\Delta_4 -2} |x-z|^{\Delta_4-2\Delta_1-2} 
\end{align}
Since the first and second lines have the pre-factors $2\Delta_1 +2-d=2\gamma_1=O(g^2)$ and $C_{1,4,1}=O(g)$, then including the $1/g$ factor, these two lines are totally the $O(g^2)$ contributions. Since we are calculating the $O(g)$ contribution to the anomalous dimensions for $\gamma_{n \ge 3}$, these terms can be neglected. On the other hand, the third line is $\frac{1}{g}\, C_{1,4,1} \Delta_4 (2 \Delta_1-\Delta_4) = O(g^0)$ and remains at the lowest-order calculation. If we notice the relation
\begin{align}
\Delta_5 =\Delta_1 +2,
\end{align}
the last term can be recast in
\begin{align}
C_{1,4,5}|x-y|^{-\Delta_1 -\Delta_4 +\Delta_5} |y-z|^{-\Delta_4-\Delta_5 +\Delta_1} |x-z|^{\Delta_4-\Delta_1-\Delta_5},
\end{align}
where we combine all the $O(g^0)$ coefficients into $C_{1,4,5}$. This is the reason why the above form of the three-point function is adequate for our purpose. For the correlation function $\braket{\phi(x) \phi^5 (y) \phi^{6} (z)  } $, the same argument can be applied. This situation is applied for the $\phi^4$-theory in $(4-\epsilon)$ dimensions. But for the $\phi^3$-theory in $(6-\epsilon)$ dimensions, the situation is quite different since the wave function renormalization start from the one-loop graph. Then for the $\phi^3$-theory, we should include the terms neglected here.

\section{$\phi^4$-theory in $(4-\epsilon)$ dimensions} 
A next example is a $\phi^4$-theory in $(4-\epsilon)$ dimensions. This can be analyzed in the same manner as the previous section, where the forms of the three-point functions including the descendant field can be approximated as if they are the correlation functions only with the primaries.

\subsubsection*{Set-up}
The action and the anomalous dimensions are defined as
\begin{align}
S&=\int d^d x \,  \left( \frac{1}{2} \partial \phi^2 +\frac{g \mu ^{ \epsilon}}{4!} \phi^4   \right),~~~d=4-\epsilon \\
\Delta_1  &:= \Delta_{\phi} =\left( 1-\frac{\epsilon}{2} \right) +\gamma_1,~~~
\Delta_n  := \Delta_{\phi^n} =n\left( 1-\frac{\epsilon}{2} \right)+\gamma_n.
\end{align}
The classical equation of motion relates $\Box \phi$ to $\phi^3$,
\begin{align}
\Box \phi =\frac{g \mu^{\epsilon} }{3!} \phi^3.
\end{align}
%

\subsubsection*{Two-point function}
We start with the two-point function of $\phi$:
\begin{align}
\braket{\phi(x)  \phi(y)} &\overset{\mathrm{tree-level}}{=} \frac{1}{4 \pi^2} \frac{1}{|x-y|^2}\\
\braket{\phi(x)  \phi(y)} &=c \, |x-y|^{-2 \Delta_1}.
\end{align}
Taking the derivatives of the above function, we have 
\begin{align}
\braket{ \Box \phi(x) \Box \phi(y)} &= c \Box_x \Box_y |x-y|^{-2 \Delta_1} \nonumber  \\
&= 2c  \Delta_1(2\Delta_1 +2) (2\Delta_1 +2-d) (2\Delta_1+4-d) |x-y|^{-2\Delta_1 -4} \nonumber  \\
&\sim \frac{8}{\pi^2} \gamma_1 |x-y|^{-6} 
\end{align}
We can evaluate the left-hand side at $O(g^2)$ by using the classical equation of motion.
\begin{align}
\braket{ \Box \phi(x) \Box \phi(y)} &= \left( \frac{g \mu^{\epsilon}}{3!} \right)^2 \braket{\phi^3(x)  \phi^3(y)}\nonumber  \\
&=\frac{g^2}{3!} \frac{1}{(4 \pi^2)^3} |x-y|^{-6}
\end{align}
Therefore we obtain
\begin{align}
\gamma_1 =\frac{1}{3 \cdot 4^3} \left( \frac{g}{4 \pi^2} \right)^2.
\end{align}
%

\subsubsection*{Three-point function}
Next, we will study the three-point functions in order to derive the anomalous dimensions of the composite operators and to find the critical value of the coupling $g_*$. Let us start with the correlator $\braket{\phi(x) \phi(y)  \phi^2(z) }$, whose general form is constrained from the conformal symmetry as
\begin{align}
\braket{\phi(x) \phi(y)  \phi^2(z)  } &=C_{1,1,2} |x-y|^{\Delta_2 -2\Delta_1} |y-z|^{-\Delta_2} |x-z|^{-\Delta_2},
\end{align}
where the tree-level coefficient of $C_{1,1,2}$ is obtained from the simple calculation of the Wick contractions and it is
\begin{align}
C_{1,1,2}^{\mathrm{tree}}=\frac{1}{8 \pi^4}.
\end{align}
First, we will take the derivative by $\Box_x$ and evaluate the right-hand side at the lowest order.
\begin{align}
\braket{\Box_x \phi(x) \phi(y)  \phi^2(z) } &= C_{1,1,2}(2\Delta_1-\Delta_2) (2 \Delta_1+2-d) |x-y|^{\Delta_2-2\Delta_1-2} |y-z|^{-\Delta_2} |x-z|^{-\Delta_2}  \nonumber \\
&\qquad +C_{1,1,2}\Delta_2 (2 \Delta_1+2-d) |x-y|^{\Delta_2-2\Delta_1} |y-z|^{-\Delta_2} |x-z|^{-\Delta_2-2}    \nonumber \\
&\qquad \quad -C_{1,1,2} \Delta_2 (2\Delta_1-\Delta_2) |x-y|^{\Delta_2-2\Delta_1-2} |y-z|^{-\Delta_2+2} |x-z|^{-\Delta_2-2}   \nonumber  \\
&=C_{1,1,2}(2\gamma_1 -\gamma_2)2\gamma_1  |x-y|^{\Delta_2-2\Delta_1-2}  |y-z|^{-\Delta_2} |x-z|^{-\Delta_2}  \nonumber \\
&\qquad  +C_{1,1,2}  \Delta_2 2\gamma_1 |x-y|^{\gamma_2-2\gamma_1} |y-z|^{-\Delta_2} |x-z|^{-\Delta_2-2}    \nonumber  \\
&\qquad \quad -C_{1,1,2} \Delta_2 (2\gamma_1-\gamma_2)   |x-y|^{\Delta_2-2\Delta_1-2} |y-z|^{-\Delta_2+2} |x-z|^{-\Delta_2-2} \nonumber  \\
& \sim  4C_{1,1,2} \gamma_1  |y-z|^{-\Delta_2} |x-z|^{-\Delta_2-2}  -4C_{1,1,2} \gamma_1 |x-y|^{-2} |x-z|^{-4}  \nonumber  \\ 
&\qquad + C_{1,1,2} \Delta_2 \gamma_2 |x-y|^{\gamma_2-2} |y-z|^{-\Delta_2+2} |x-z|^{-\Delta_2-2} +O(g^3)\nonumber  \\
&\sim 2C_{1,1,2} \gamma_2 |x-y|^{-2}|x-z|^{-4} +O(g^2),
\end{align}
where we used the fact that $\gamma_1$ starts with $O(g^2)$, which is the result from the study of the two-point function.
By using the classical equation of motion, the left-hand side becomes
\begin{align}
\braket{\Box_x \phi(x) \phi(y)  \phi^2(z) } &=  \frac{g}{3!}  \braket{\phi^3 (x) \phi (y) \phi^2 (z)} \nonumber \\
&\sim \frac{g}{ (4 \pi^2)^3} |x-y|^{-2}  |x-z|^{-4} +O(g^2)
\end{align}
Then, comparing these two, we find
\begin{align}
\gamma_2=\frac{g}{16 \pi^2}. \label{g2}
\end{align}
One may expect that if we apply further the derivative $\Box_y$ to the above three-point function, we get the $O(g^2)$ result of the anomalous dimension $\gamma_2$. However this is not the case since we will get the uninteresting relation
\begin{align}
\gamma_2^2 =O(g^2) .
\end{align}

For the composite operators $\phi^n$, we study the following three-point functions.
\begin{align}
\braket{\phi(x) \phi^n (y) \phi^{n+1}(z)} =C_{1,n,n+1}|x-y|^{\Delta_{n+1} - \Delta_{1}-\Delta_{n}} |y-z|^{\Delta_{1}-\Delta_{n}-\Delta_{n+1}} |x-z|^{\Delta_{n}-\Delta_{1}-\Delta_{n+1}}
\end{align}
For $n=2,3$, the correlator includes a descendant field $\phi^3$, then the right-hand side generally involves other terms. However the coefficients of such terms start with $O(g)$ or more higher power of $g$. In the following we will consider the $O(g)$ contribution, so these terms can be neglected as before. Notice that we take a derivative of this correlator and this manipulation gives rise to another $O(g)$ pre-factor, so in the above we have to only retain the terms with $O(g^0)$ coefficients. The tree-level coefficient $C_{1,n,n+1}$ is given by    
\begin{align}
C_{1,n,n+1}^{\mathrm{tree}}=\frac{(n+1)!}{(4\pi^2)^{n+1}}.
\end{align}
As mentioned above, we will take a derivative by $\Box_x$ and compare it with the result from the classical equation of motion.
\begin{align}
\braket{\Box_x \phi(x) \phi^n (y) \phi^{n+1}(z)} &= 2C_{1,n,n+1}(\gamma_{n+1}-\gamma_n) |x-y|^{-2} |y-z|^{-2n+2} |x-z|^{-4} +O(g^2)\\
\braket{\Box_x \phi(x) \phi^n (y) \phi^{n+1}(z)} &=\frac{g}{3!}\braket{ \phi^3(x) \phi^n (y) \phi^{n+1}(z)} \nonumber \\
&\sim g\, n! \, _{n+1}C_2 \frac{1}{(4 \pi^2)^{n+2}} |x-y|^{-2} |y-z|^{-2n+2} |x-z|^{-4} 
\end{align}
Then we find the following recurrence relation
\begin{align}
\gamma_{n+1} -\gamma_n =\frac{g}{4 \pi^2} \frac{n}{4}.
\end{align}
Remembering the result \eqref{g2}, $\gamma_2=\frac{g}{16 \pi^2}$, this is easily solved as
\begin{align}
\gamma_n=\frac{g}{32 \pi^2}n(n-1),~~~\mbox{for}~~ n \ge 2.
\end{align}
From the assumption $\Delta_3 = \Delta_1+2 $, we have
\begin{align}
\gamma_3-\gamma_1=\epsilon
\end{align}
and the critical coupling is found to be
\begin{align}
g_*=\frac{16 \pi^2}{3} \epsilon.
\end{align}
The anomalous dimensions of the composite operators are summarized as
\begin{align}
\gamma_1=\frac{\epsilon^2}{108} ,~~~\gamma_{n \ge 2} =\frac{1}{6}n(n-1)\epsilon +O(\epsilon^2 ), 
\end{align}
which is consistent with the perturbative results \cite{Kleinert:2001ax, Kehrein:1992fn} and the conformal method developed in \cite{Rychkov:2015naa}.

\section{$\phi^3$-theory in $(6-\epsilon)$ dimensions} 
In this section we study the $\phi^3$-theory in $(6-\epsilon)$ dimensions. As claimed earlier, in this case we are very careful about the three-point functions including the $\phi^2$ fields since $\phi^2$ is a descendant and the three-point functions become more involved.
The bare Lagrangian is
\begin{align}
S=\int d^d x \,  \left( \frac{1}{2} \partial \phi^2 +\frac{g \mu ^{ \epsilon/2}}{3!} \phi^3   \right),~~~d=6-\epsilon
\end{align}
and the scaling dimensions are defined as
\begin{align}
\Delta_1  &:= \Delta_{\phi} =\left( 2-\frac{\epsilon}{2} \right) +\gamma_1,~~~\Delta_n  := \Delta_{\phi^n} =n\left( 2-\frac{\epsilon}{2} \right)+\gamma_n.
\end{align}
%
The classical equation of motion is
\begin{align}
\Box \phi =\frac{g \mu^{\epsilon/2} }{2} \phi^2.
\end{align}
%
\subsubsection*{Two-point function}
For two-point functions, the manipulation is completely the same as the previous two sections.
\begin{align}
\braket{\phi(x)  \phi(y)} &\overset{\mathrm{tree-level}}{=} \frac{1}{4 \pi^3} \frac{1}{|x-y|^4}\\
\braket{\phi(x)  \phi(y)} &=c \, |x-y|^{-2 \Delta_1}
\end{align}
For the case where a single d'Alembertian $\Box_x$ acts on the above two-point function, the resulting correlation function $\braket{\phi^2 (x)  \phi(y) }$ is vanishing at the tree-level approximation. Then we need to take the derivatives of the two-point function by $\Box_x$ and $\Box_y$,
\begin{align}
\braket{ \Box \phi(x) \Box \phi(y)} &= c \Box_x \Box_y |x-y|^{-2 \Delta_1} \nonumber  \\
&= 2c  \Delta_1(2\Delta_1 +2) (2\Delta_1 +2-d) (2\Delta_1+4-d) |x-y|^{-2\Delta_1 -4} \nonumber \\
&\sim 4^2 \cdot 6 c \gamma_1 |x-y|^{-2\Delta_1 -4} \nonumber \\
&\sim \frac{24}{\pi^3} \gamma_1 |x-y|^{-8}, \\
\braket{ \Box \phi(x) \Box \phi(y)} &= \left( \frac{g \mu^{\epsilon/2}}{2} \right)^2 \braket{\phi^2(x)  \phi^2(y)} \nonumber \\
&=\frac{g^2}{2} \frac{1}{(4 \pi^3)^2} |x-y|^{-8}
\end{align}
Therefore we find that the $O(g)$ contribution to $\gamma_1$ is absent and
\begin{align}
\gamma_1 =\frac{1}{3 \cdot 4^3} \frac{g^2}{4 \pi^3}. 
\end{align}

\subsubsection*{Three-point function}
First, let us assume that the field $\phi^n$ is a conformal primary with the scaling dimension $\Delta_n$ except for $n=2$. The three-point function of $\phi$ is fixed by the conformal symmetry as 
\begin{align}
\braket{\phi(x) \phi(y) \phi(z)} &=C_{1,1,1} |x-y|^{-\Delta_1} |y-z|^{-\Delta_1} |x-z|^{-\Delta_1}. 
\end{align}
Since the above three-point function is zero for the free theory limit, $C_{1,1,1}=O(g)$.
By taking a derivative by $\Box_x$, we obtain
\begin{align}
\braket{\Box_x \phi(x) \phi(y) \phi(z)} &=C_{1,1,1} \Delta_1 (2\Delta_1 +2-d) |x-y|^{-\Delta_1-2} |y-z|^{-\Delta_1} |x-z|^{-\Delta_1} \nonumber \\
&\qquad + C_{1,1,1} \Delta_1 (2\Delta_1 +2-d) |x-y|^{-\Delta_1} |y-z|^{-\Delta_1} |x-z|^{-\Delta_1-2} \nonumber \\ 
&\qquad \quad - C_{1,1,1} (\Delta_1)^2 |x-y|^{-\Delta_1-2} |y-z|^{-\Delta_1+2} |x-z|^{-\Delta_1-2}\nonumber  \\
&\sim -4C_{1,1,1} |x-y|^{-4} |x-z|^{-4}+O(g^2).
\end{align}
Using the classical equation of motion, we find
\begin{align}
\braket{\Box_x \phi(x) \phi(y) \phi(z)} &=\frac{g}{2} \braket{\phi(x)^2 \phi(y) \phi(z)} \nonumber \\
&\sim \frac{g}{(4 \pi^3)^2} |x-y|^{-4} |x-z|^{-4}+O(g^2),
\end{align}
and then
\begin{align}
C_{1,1,1}=-\frac{g}{64 \pi^6}+O(g^2).
\end{align}
Next, we will consider the three-point function including the descendant $\phi^2$. This can be obtained from the three-point function $\braket{\phi(x) \phi(y) \phi(z)}$ since we are now assuming $\phi^2$ is a descendant. At the leading order we find
\begin{align}
\braket{\phi(x) \phi(y) \phi^2(z)} &=  \frac{2}{g} \braket{\phi(x) \phi(y)  \Box_z \phi(z)} \nonumber \\
&\sim -\frac{\gamma_1}{8 \pi^6} |x-y|^{-2} |y-z|^{-4} |x-z|^{-2} -\frac{\gamma_1}{8 \pi^6}|x-y|^{-2} |y-z|^{-2} |x-z|^{-4} \nonumber \\
&\qquad + \frac{1}{8 \pi^6} |x-y|^{-\Delta_1+2}|y-z|^{-\Delta_1-2} |x-z|^{- \Delta_1-2},
\end{align}
where we omitted the unnecessary terms and factors for our purpose of the $O(g^2)$ computation. For example, in the second line, the terms are proportional to $\gamma_1$, so the powers of $|x-y|, |y-z|$ and $|x-z|$ can be replaced with the classical values. Notice that the difference between this expression and the 3d or 4d ones. In the previous two sections, we can drop the first two terms since the anomalous dimension $\gamma_1$ started with $O(g^2)$ and the one of the descendant field starts with $O(g)$. However, in the six-dimensional $\phi^3$-theory, these two quantities would start with the same order, then we should retain all the three terms above. 

Being multiplied by $\Box_x$ and $\Box_y$, the right-hand side becomes
\begin{align}
\braket{ \Box_x \phi(x)  \Box_y \phi(y) \phi^2(z)} &= \left(\frac{4 \gamma_1}{\pi^6} +\frac{(2\gamma_1-\epsilon)}{\pi^6} \right)|x-y|^{-4} |y-z |^{-4}|x-z|^{-4}.
\end{align}
On the other hand, applying the classical equation of motion, we find
\begin{align}
\braket{ \Box_x \phi(x)  \Box_y \phi(y) \phi^2(z)} &=\frac{g^2}{4} \braket{\phi^2(x) \phi^2(y) \phi^2(z)} \nonumber \\
&\sim \frac{g^2 2^3}{4(4 \pi^3)^3}|x-y|^{-4} |y-z |^{-4}|x-z|^{-4}.
\end{align}
Comparing these two results, we obtain
\begin{align}
\gamma_1=\frac{\epsilon}{6} +\frac{g^2}{3 \cdot 4^3 \pi^3}.
\end{align}
Combining this with
\begin{align}
\gamma_1 =\frac{1}{3 \cdot 4^3} \frac{g^2}{4 \pi^3}, 
\end{align}
we find the critical coupling and the critical exponent
\begin{align}
g_*^2 =-\frac{2 \cdot 4^3 \pi^3 }{3} \epsilon +O(\epsilon^2),~~~\gamma_1(g^2_*) =-\frac{1}{18} \epsilon+O(\epsilon^2).
\end{align}
This result is consistent with the perturbation results (see, for example, \cite{Fisher:1978pf,deAlcantaraBonfim:1980pe,deAlcantaraBonfim:1981sy}). The conformal method developed in \cite{Rychkov:2015naa} was not applicable to this theory since they crucially used the conditions that $\gamma_1=O(\epsilon^2)$ and $\gamma_{n>1} =O(\epsilon)$. However our method does not rely on these condition and can be widely applied.

\section{$\phi_i \phi_i \sigma $ interaction in $(6-\epsilon)$ dimensions} 
This section generalizes the $\phi^3$-theory in $d=(6-\epsilon)$ dimensions by adding the $\phi_i \phi_i \chi $ interaction. We will also include the $\chi^3$ term.
\begin{align}
S=\int d^dx \, \frac{1}{2} (\partial \phi_i)^2 +\frac{1}{2} (\partial \chi)^2 +\frac{g_1}{2} \chi \phi_i \phi_i +\frac{g_2}{3!} \chi^3 ,~~~i=1,\cdots ,N
\end{align}
The theory is recently studied in \cite{Fei:2014yja,Giombi:2016hkj} (for an old reference, see for example, \cite{Mikhailov:1985cm}).
The $\beta$ function of $g_1$ is very similar to the one of the four-dimensional QCD. When we limit to the case with $g_2=0$ and $\epsilon =0$, the theory is asymptotically free for $N<8$ and we have the IR free phase for $N >8$, . For $N=8$, the $\beta$ function is vanishing at least from the one-loop analysis and it would show the conformal symmetry and we can use the conformal method. 
The theory can also have the Lee-Yang zero for the imaginary values of the couplings. For the large $N$ limit, the Wilson-Fisher fixed point is realized at the real values of the couplings.
The classical equations of motion for $\phi_i$ and $\chi$ are
\begin{align}
\Box \phi_i &=g_1 \chi \phi_i,~~~~~\Box \chi =\frac{g_1}{2} \phi_k \phi_k +\frac{g_2}{2} \chi^2
\end{align}
The scaling dimensions are defined as
\begin{align}
\Delta_{\phi} &=(2 -\frac{\epsilon}{2}) +\gamma_\phi,~~~\Delta_{\chi} ~=(2 -\frac{\epsilon}{2}) +\gamma_\chi  
\end{align}
%
\subsubsection*{Two-point function}
The theory includes the two types of the scalar fields, $\phi_i$ and $\chi$, then we have the two propagators. 
\begin{align}
\braket{\phi_i(x)  \phi_j (y)} &=c_{\phi} \delta_{ij} |x-y|^{-2 \Delta_{\phi}} \\
\braket{\chi (x)  \chi(y) } &=c_{\chi}  |x-y|^{-2 \Delta_{\chi}}
\end{align}
Applying the two differential operators, $\Box_x$ and $\Box_y$, we obtain the following result.
\begin{align}
\gamma_\phi &=\frac{1}{6} \frac{g_1^2 }{(4 \pi)^3} \label{gammaphi} \\
\gamma_\chi &= \frac{1}{12} \left( \frac{g_1^2 N}{(4 \pi)^3} +\frac{g_2^2}{(4 \pi)^3} \right), \label{gammachi}
\end{align}
where we used the classical equations of motion. This is precisely identical to the one-loop calculation.

\subsubsection*{Three-point function}
Next, let us study the three-point functions which effectively determine the $\beta$ functions at the leading order and the critical values of the couplings.
The analysis of the three-point function is more subtle since now the theory has the two types of the scalar fields and the equations of motion should involves the various terms. 

We start with the correlation function, $\braket{ \phi_i (x)  \phi_j (y) \chi (z)}$. By the assumption that the fields $\phi_i$ and $\chi$ are the primary fields, the coordinate dependence is completely fixed as
\begin{align}
\braket{ \phi_i (x)  \phi_j (y) \chi (z)} =C_{ij \chi} |x-y|^{\Delta_{\chi} -2\Delta_{\phi}} |y-z|^{-\Delta_{\chi}} |x-z|^{-\Delta_{\chi}},
\end{align}
where the three-point function above is classically vanishing, then the coefficient is $O(g_1)$. We will first determine the leading behavior of $C_{ij \chi}$.  We multiply the above function by the differential operator $\Box_z$, we find
\begin{align}
\braket{ \phi_i (x)  \phi_j (y) \Box_z \chi (z)} &=C_{ij \chi} \Box_z |x-y|^{\Delta_{\chi} -2\Delta_{\phi}} |y-z|^{-\Delta_{\chi}} |x-z|^{-\Delta_{\chi}}  \nonumber \\
&\sim -4 C_{ij \chi} |y-z|^{-4} |x-z|^{-4} \\
\braket{ \phi_i (x)  \phi_j (y) \Box_z \chi (z)} &=\braket{\phi_i (x)  \phi_j (y)  \left( \frac{g_1}{2} \phi_k \phi_k +\frac{g_2}{2} \chi^2  \right)}  \nonumber \\
&\sim \frac{g_1 \delta_{ij}}{(4 \pi^3)^2} |y-z|^{-4}|x-z|^{-4},
\end{align}
where in the third line above, we used the classical equation of motion for $\chi$. Comparing the second and fourth lines, the coefficient is determined to be 
\begin{align}
\left. C_{ij \chi} \right|_{\mathrm{lowest}} =-\frac{g_1}{64 \pi^6} \delta_{ij} 
\end{align}
at the leading order of the perturbation. The similar manipulation for the three-point function $\braket{ \chi (x)  \chi (y) \chi (z)} $ gives
\begin{align}
\left. C_{\chi \chi \chi}  \right|_{\mathrm{lowest}}  =-\frac{g_2}{64 \pi^6}.
\end{align}

Next we will study the correlation functions which involve the descendant field. The form of the correlation function is fixed by the equation of motion and the three-point function of the primary fields.
\begin{align}
\braket{ \phi_i (x)  \phi_j (y) (\phi_k \phi_k) (z)} &= \frac{2}{g_1} \braket{ \phi_i (x)  \phi_j (y) (\Box \chi -\frac{g_2}{2} \chi^2 ) (z)}  \nonumber \\
&=  \frac{2}{g_1}\Box_z \braket{\phi_i (x)  \phi_j (y) \chi(z)} -\frac{g_2}{g_1} \braket{ \phi_i (x)  \phi_j (y) \chi^2 (z)} 
\end{align}
We multiply the above expression by the derivatives $\Box_x$ and $\Box_y$,
\begin{align}
\braket{ \Box \phi_i (x) \Box \phi_j (y) (\phi_k \phi_k) (z)} &= \frac{2}{g_1}C_{ij\chi} \Box_x \Box_y \Box_z  |x-y|^{\Delta_{\chi} -2\Delta_{\phi}}|y-z|^{-\Delta_{\chi}} |x-z|^{-\Delta_{\chi}} \nonumber \\ 
& \qquad \quad -\frac{g_2}{g_1} \braket{ \Box\phi_i (x)\Box  \phi_j (y) \chi^2 (z)} \label{pppp}
\end{align}
The left-hand side at the leading order can be evaluated by the classical equation of motion as
\begin{align}
\braket{ \Box \phi_i (x) \Box \phi_j (y) (\phi_k \phi_k) (z)}
&=g_1^2\braket{ \chi \phi_i(x)  \chi \phi_j (y)  (\phi_k \phi_k) (z) }  \nonumber \\
&\sim \frac{2g_1^2 \delta_{ij}}{64 \pi^9} |x-y|^{-4}|y-z|^{-4} |x-z|^{-4} 
\end{align}
The right-hand side in \eqref{pppp} is also evaluated as foliows.
%
\begin{align}
& \frac{2}{g_1}C_{ij\chi} \Box_x \Box_y \Box_z  |x-y|^{\Delta_{\chi} -2\Delta_{\phi}}|y-z|^{-\Delta_{\chi}} |x-z|^{-\Delta_{\chi}} \nonumber \\
 & \qquad \qquad \sim \frac{2 \delta_{ij}}{\pi^6} (2 \gamma_\phi +\gamma_\chi -\frac{\epsilon}{2}) |x-y|^{-4}|y-z|^{-4} |x-z|^{-4} 
\end{align}
\begin{align}
-\frac{g_2}{g_1} \braket{ \Box\phi_i (x)\Box  \phi_j (y) \chi^2 (z)} &=- g_1g_2  \braket{ \chi \phi_i (x)  \chi \phi_i (y) \chi^2 (z) }   \nonumber \\
& \sim -\frac{2g_1g_2 \delta_{ij}}{64 \pi^9} |x-y|^{-4}|y-z|^{-4} |x-z|^{-4} 
\end{align}
Comparing the both sides we find
\begin{align}
2\gamma_\phi +\gamma_\chi =\frac{\epsilon}{2} + \frac{g_1^2}{64 \pi^3} +\frac{g_1g_2}{64\pi^3}.
\end{align}
If we substitute the results \eqref{gammaphi} and \eqref{gammachi} in the above, we find
\begin{align}
-\frac{\epsilon}{2} +\frac{1}{(4\pi)^3} \frac{(N-8)g_1^2 -12g_1g_2 +g_2^2}{12} =0
\end{align}
This is precisely the same as the $\beta$ function for $g_1$,
\begin{align}
\beta_1 =-\frac{\epsilon}{2}g_1 +\frac{1}{(4\pi)^3} \frac{(N-8) g_1^3 -12g_1^2g_2 +g_1g_2^2 }{12}
\end{align}
and it is consistent with the fact that for $g_2=0$, $N=8$ and $\epsilon=0$, the theory is conformal at the leading order of the perturbation.

Next, we would like to find the $\beta$ function for $g_2$. Then it is plausible to study the correlation functions, $\braket{\chi(x) \chi(y) \chi^2 (z)} $ and $\braket{\chi(x) \chi(y) \chi (z)} $. Since $\chi^2$ is related to $\Box \chi$ and $\phi_k \phi_k$ by the equation of motion, we have the following relation.
\begin{align}
\braket{\chi(x) \chi(y) \chi^2 (z)} &=\frac{2}{g_2} \braket{\chi(x) \chi(y)  \left( \Box_z \chi -\frac{g_1}{2} \phi_k \phi_k \right)(z)}\\
\braket{\Box_x\chi(x) \Box_y\chi(y) \chi^2 (z)} &=\frac{2}{g_2}\Box_x \Box_y \Box_z \braket{\chi(x) \chi(y) \chi (z)} -\frac{g_1}{g_2} \braket{\Box_x \chi(x) \Box_y \chi(y)  \phi_k \phi_k (z) } \label{ccc}
\end{align}
The left-hand side in \eqref{ccc} can be evaluated as
\begin{align}
\braket{\Box_x\chi(x) \Box_y\chi(y) \chi^2 (z)} &= \braket{\left( \frac{g_1}{2} \phi_k \phi_k +\frac{g_2}{2} \chi^2 \right)(x) \left( \frac{g_1}{2} \phi_k \phi_k +\frac{g_2}{2} \chi^2 \right)(y) \chi^2 (z)} \nonumber \\
&\sim \frac{g_2^2}{4}\braket{\chi^2 (x) \chi^2(y) \chi^2(z)} \nonumber \\
&\sim \frac{2g_2^2}{ 64 \pi^9}  |x-y|^{-4}|y-z|^{-4} |x-z|^{-4}. 
\end{align}
The two terms in the right-hand side of \eqref{ccc} become
\begin{align}
\frac{2}{g_2}\Box_x \Box_y \Box_z \braket{\chi(x) \chi(y) \chi (z)} &= \frac{2}{\pi^6} (3 \gamma_\chi -\frac{\epsilon}{2})  |x-y|^{-4}|y-z|^{-4} |x-z|^{-4} \\
-\frac{g_1}{g_2} \braket{\Box \chi(x) \Box \chi(y)  \phi_k \phi_k (z) } &=-\frac{2Ng_1^3}{g_2} |x-y|^{-4}|y-z|^{-4} |x-z|^{-4},
\end{align}
where we used the general form of the three-point function $\braket{\chi(x) \chi(y) \chi (z)}$, which can be constrained by the conformal symmetries by the assumption that $\chi$ is a primary field. Comparing the both sides we obtain
\begin{align}
\frac{2g_2^2}{64 \pi^3} =-\frac{2Ng_1^3/g_2 }{64 \pi^3} +\frac{2}{\pi^6} (3 \gamma_\chi -\frac{\epsilon}{2}),
\end{align}
which is equivalent to 
\begin{align}
-\frac{\epsilon}{2} +\frac{1}{(4\pi)^3} \frac{-4N g_1^3 +Ng_1^2g_2 -3 g_2^3}{4g_2} =0.
\end{align}
Again, this is consistent with the $\beta$ function for $g_2$
\begin{align}
\beta_2 =-\frac{\epsilon}{2}g_2  +\frac{-4N g_1^3 +Ng_1^2g_2 -3 g_2^3}{4 (4 \pi)^3}.
\end{align}
In this way we correctly reproduce the anomalous dimensions for $\phi_i$ and $\chi$ and the conditions that the $\beta$ functions vanish. It seems difficult to study the composite operators $\phi_k \phi_k$ and $\chi^2$ since these two operators mix with each other and solving the mixing will inevitably require the one-loop computation. It would be interesting to calculate the mixing matrix from the tree-level calculation and this direction will be left as a future direction.

\section{Summary and Discussion}
Motivated by the recent paper \cite{Rychkov:2015naa}, we here derived the leading expression of the anomalous dimensions of the operators $\phi^n$ for the $\phi^6$-, $\phi^4$-, and $\phi^3$-theories in $(3-\epsilon), (4-\epsilon)$ and $(6-\epsilon)$ dimensions respectively. The method developed here does not rely on the Feynman diagrammatic technique but on the conformal symmetry and on the classical equations of motion. This is very parallel and similar to the method developed in \cite{Rychkov:2015naa}. 

One of the differences is that in \cite{Rychkov:2015naa}, the operator relation, $\Box  \phi =\alpha \phi^3$ is not the same as the equation of motion but just an operator identity coming from the assumption that $\phi^3$ is a descendant of the conformal primary field $\phi$ at the Wilson-Fisher fixed point. In this sense, the method in \cite{Rychkov:2015naa} does not rely on the perturbative approach. On the other hand, in this paper, we used the classical equation of motion, which is just the lowest order form of the Schwinger-Dyson equation without the contact term and clearly relying on the Lagrangian-based (perturbative) approach. 

The other difference is that in \cite{Rychkov:2015naa} they studied the three-point functions, $\braket{\phi^n(x) \phi^{n+1}(y)\phi(z) } $ and $\braket{\phi^n(x) \phi^{n+1}(y)\phi^3(z) } $, by using the OPE between $\phi^n$ and $\phi^{n+1}$. Then this is effectively equivalent to studying the three-point functions, $\braket{\phi^n(x) \phi^{n+1}(y)\phi(z) } $ and $\braket{\phi^n(x) \phi^{n+1}(y) \Box \phi(z) } $, in our approach. However, as we have mentioned in Section 4, it is important to study the three-point functions, $\braket{\phi(x) \phi(y)\phi^2(z) }$ and $\braket{\Box \phi(x)  \Box\phi(y)\phi^2(z) }$ in the $\phi^3$-theory in $(6-\epsilon)$ dimensions. Since the later one includes the two differential operators, the information on $\braket{\Box \phi(x)  \Box\phi(y)\phi^2(z) }$ is not taken into account in the method \cite{Rychkov:2015naa}. This is the reason why the method \cite{Rychkov:2015naa} can not be directly applied to the six-dimensional $\phi^3$-theory.

In this paper, we found the critical coupling $g_*=g_*(\epsilon)$ at the leading order without any perturbative input. This can be carried out by considering the two- and three-point functions and their derivatives. The coordinate dependence of these functions is constrained from the conformal symmetry. By combining these results with the classical equations of motion we obtained the anomalous dimensions and the critical coupling. It would be interesting to go beyond the leading order approximation. For example, if we want to know the next-leading order, we will need to do the one-loop computation since the use of the equation of motion reduces the loop order required for calculating the physical quantities. Thus we must inevitably deal with the regularization and the renormalization at this order.

In Section 5, we study the theory with the $N+1$ scalar fields,
\begin{align}
S=\int d^6x \, \frac{1}{2} (\partial \phi_i)^2 +\frac{1}{2} (\partial \chi)^2 +\frac{g_1}{2} \chi \phi_i \phi_i  +\frac{g_2}{3!} \chi^3,~~i=1,\cdots ,N
\end{align}
and we found the anomalous dimensions of the operators $\phi_i$ and $\chi$. For the composite operators, we have to be careful with the operator mixing, for example, between $\phi_i\phi_i$ and $\chi^2$. In our approach it seems difficult to solve the mixing only from the tree-level calculation. It would be interesting to find the techniques to avoid the one-loop computation.

It would be important to apply the method to more complicated theories including the various kinds of the scalar and fermion fields \cite{Raju:2015fza,Ghosh:2015opa} and also important to know if the conformal techniques developed here can be applied to the gauge theories.

\section*{Acknowledgments}
I would like to thank Slava Rychkov, Masashi Hayakawa, Dileep Jatkar and Yu Nakayama for valuable comments and helpful discussions.

\bibliographystyle{ieeetr}
\bibliography{multiplet}

\end{document}